\documentclass[traditabstract]{aa} 

\usepackage{graphicx,longtable,cancel}
\usepackage{txfonts}
\usepackage{natbib}
\usepackage{rotating}
\usepackage{longtable}
\usepackage{supertabular}
\usepackage{float,lscape}
\begin{document}
\bibliographystyle{aa}

\makeatletter
\def\@biblabel#1{\hspace*{-\labelsep}}
\makeatother

\title{XMMSL1J063045.9-603110: \\ a tidal disruption event fallen into the back burner}

\author{
Deborah Mainetti$\rm ^{1, 2, 3}$, Sergio Campana$\rm ^{2}$ and Monica Colpi$\rm ^{1, 3}$ 
}
\institute{
Dipartimento di Fisica G. Occhialini, Universit\`a degli Studi di Milano Bicocca, Piazza della Scienza 3, I-20126 Milano, Italy
\and
INAF, Osservatorio Astronomico di Brera, Via E. Bianchi 46, I-23807 Merate (LC), Italy
\and
INFN, Sezione di Milano-Bicocca, Piazza della Scienza 3, I-20126 Milano, Italy
}
              \date{\email{d.mainetti1@campus.unimib.it}}
              
              \abstract{Black holes at the centre of quiescent galaxies can be switched on when they accrete gas that is gained from stellar tidal disruptions. A star approaching a black hole on a low angular momentum orbit may be ripped apart by tidal forces, which triggers raining down of a fraction of stellar debris onto the compact object through an accretion disc and powers a bright flare. In this paper we discuss XMMSL1J063045.9-603110 as a candidate object for a tidal disruption event. The source has recently been detected to be bright in the soft X-rays during an XMM-Newton slew and later showed an X-ray flux decay by a factor of about 10 in twenty days. We analyse XMM-Newton and Swift data. XMMSL1J063045.9-603110 shows several features typical of tidal disruption events: the X-ray spectrum shows the characteristics of a spectrum arising from a thermal accretion disc, the flux decay follows a $t^{-5/3}$ law, and the flux variation is $> 350$. Optical observations testify that XMMSL1J063045.9-603110 is probably associated with an extremely small galaxy or even a globular cluster, which suggests that intermediate-mass black holes are located in the cores of (at least) some of them.}
              
              
               \authorrunning{Mainetti et al.} 
\maketitle

\section{Introduction}
The formation and evolution processes of supermassive black holes (SMBHs), which live in the centre of massive galaxies (Kormendy \& Richstone 1995; Kormendy \& Gebhardt 2001), are one of the main riddles in astrophysics today. It is commonly accepted that when these SMBHs are settled at the centre of their host galaxies, they grow mainly by accretion of the surrounding gas and merging with smaller black holes (Soltan 1982; Yu \& Tremaine 2002). Major inflows of gas drive the emission of huge amounts of energy that power active galactic nuclei (AGNs), even though AGNs typically have short duty cycles and galactic central SMBHs are mostly in a low luminous state (Ho 2008).
The lighting engine of a quiescent SMBH can also be the tidal disruption (TD) of a star orbiting it (Rees 1988). Dynamical encounters in the nuclear star cluster increase the probability for a star to be scattered close to the SMBH on a low angular momentum orbit, in which it would experience the SMBH tidal field (Alexander 2012). When the stellar self-gravity is no longer able to counteract the SMBH tidal force, the star is disrupted. A fraction of the resulting stellar debris is bound to the SMBH and accretes onto it through an accretion disc. This triggers a peculiar flaring emission (e.g. Rees 1988; Phinney 1989).     

The critical pericentre distance of a star for TD is the black hole (BH) tidal radius 
\begin{equation}
r_{\rm t}\sim R_{\rm *}\Bigl(\frac{M_{\rm BH}}{M_{\rm *}} \Bigr)^{1/3}\sim 10^2 \rm R_{\rm \odot} \Bigl(\frac{\it R_{\rm *}}{1 \rm R_{\rm \odot}}\Bigr)\Bigl(\frac{\it M_{\rm BH}}{10^6 \rm M_{\rm \odot}}\Bigr)^{1/3}\Bigl(\frac{1 \rm M_{\rm \odot}}{\it M_{\rm *}}\Bigr)^{1/3}		\label{rtidal},
\end{equation}
with $R_{\rm *}$ and $M_{\rm *}$ being the star radius and mass and $M_{\rm BH}$ the BH mass (Hills 1975; Frank \& Rees 1976). If 
\begin{equation}
r_{\rm t}<r_{\rm s},		\label{rtrs}
\end{equation}
where
\begin{equation}
r_{\rm s}= \frac{\rm x\it GM_{\rm BH}}{c^2}\sim 5\rm R_{\rm \odot} \Bigl(\frac{\it M_{\rm BH}}{10^6 \rm M_{\rm \odot}}\Bigr)\Bigl(\frac{x}{2}\Bigr)	\label{rschwarz}
\end{equation}
is the BH event horizon radius (x=2 for non-rotating BHs), the star enters the BH horizon before it is tidally disrupted and no  flares are observable. For non-rotating BHs (Kesden 2012), Equation \ref{rtrs} implies that solar-type stars, white dwarfs, and giant stars are swallowed entirely if the $M_{\rm BH}$ is greater than $10^8 \rm M_{\rm \odot}$, $10^5 \rm M_{\rm \odot}$ and $10^{10} \rm M_{\rm \odot}$, respectively. In contrast, less massive BHs in quiescent (or low-luminous) galaxies can be inferred from the observation of tidal disruption flares. 
TDs are observationally estimated to occur at a rate of about $10^{-5} \rm galaxy^{-1} \rm yr^{-1}$ (Donley et al. 2002). Although they are rare and observations are sparse, several candidate TDs have been discovered in the optical-UV (Gezari 2012 and references therein) and soft X-ray bands (Komossa 2012; Komossa 2015 and references therein), where the peak of an accretion disc emission lies (Strubbe \& Quataert 2009), and in the hard X-ray-radio band, where they are accompanied by a jetted emission (e.g. Bloom et al. 2011; Burrows et al. 2011; Cenko et al. 2012; Hryniewicz \& Walter 2016). The BH mass can sometimes be estimated for candidate TDs. 

Tidal disruption candidates have recently also been reported in dwarf galaxies, which suggests that an intermediate-mass BH (IMBH; $10^2 \rm M_{\rm \odot} \leq \it M_{\rm BH} \leq \rm 10^6 \rm M_{\rm \odot}$; Ghosh et al. 2006; Maksym et al. 2013; Maksym et al. 2014a; Donato et al. 2014; Maksym et al. 2014b) is located in their nuclei. The formation process of IMBHs is still an open question (e.g. Madau \& Rees 2001; Miller \& Hamilton 2002; Portegies Zwart \& McMillan 2002; Begelman et al. 2006; Latif et al. 2013), but confirming their existence, detecting them, and obtaining a mass estimate are extremely important, as they could fill in the current gap in mass distribution between stellar-mass BHs and SMBHs (Merloni \& Heinz 2013) and also explain the origin of SMBHs through mergers of small galaxies hosting IMBHs (e.g. Volonteri 2010). TDs in dwarf galaxies are an opportunity for achieving all this.             

In this paper we discuss XMMSL1J063045.9-603110, a recently discovered bright soft X-ray source whose X-ray activity might be attributable to a TD.  
We briefly describe the fundamental structure of TDs (Sect. \ref{fundamentals}) and summarise what is known about XMMSL1J063045.9-603110 (Sect. \ref{known}). We discuss the possible TD nature of the source in an extremely small galaxy or even in a globular cluster, reducing (Sect. \ref{data_reduction}) and exploring X-ray data from spectral analysis (Sect. \ref{data_analysis}) to flux variability (Sect. \ref{variability}) and also investigate the activity of XMMSL1J063045.9-603110 at lower energies (Sect. \ref{UVOT}) and evaluate the probably host absolute magnitude (Sect. \ref{cTD}).
Our results are summarised in Sect. \ref{conclusions}.

\section{Tidal disruption} \label{fundamentals}
A star approaching the central BH of a galaxy on a low angular momentum orbit may be swallowed whole by the compact object or be tidally disrupted outside the BH horizon. The latter event occurs if the BH tidal radius $r_{\rm t}$ (Eq. \ref{rtidal}) is greater than the BH event horizon radius $r_{\rm s}$ (Eq. \ref{rschwarz}), that is, if 
\begin{equation}
M_{\rm BH}< \frac{c^3}{(\rm x \it G)^{\rm 3/2}}\frac{R_{\rm *}^{3/2}}{M_{\rm *}^{1/2}}\sim 10^8 \rm M_{\rm \odot} \Bigl(\frac{\it R_{\rm *}}{1 \rm R_{\rm \odot}}\Bigr)^{3/2}\Bigl(\frac{1 \rm M_{\rm \odot}}{\it M_{\rm *}}\Bigr)^{1/2}\Bigl(\frac{2}{\rm x}\Bigr)^{3/2},		\label{tdfyes}
\end{equation}
when the BH tidal force overcomes the star self-gravity, which prevents the star from remaining assembled.
In this regime, the star is completely disrupted if its pericentre distance $r_{\rm p}$ is shorter than about $r_{\rm t}$.\footnote{For $r_{\rm p}\gtrsim r_{\rm t}$ the star is only partially disrupted (MacLeod et al. 2012; Guillochon \& Ramirez-Ruiz 2013, 2015a). We do not describe this case here in detail because the physics of partial TDs recovers that of total TDs to a first approximation.} When the star is on a parabolic orbit, nearly half of the resulting stellar debris is scattered onto highly eccentric orbits bound to the BH with a spread in the specific orbital energies of $\Delta E\sim GM_{\rm BH}R_{\rm *}/r_{\rm t}^2$
(Lacy et al. 1982). The other half of stellar debris leaves the system on hyperbolic orbits.
The first returning time at pericentre of the bound debris depends on their new orbital energy through Kepler's third law, and for the most bound material it is 
\begin{equation}
t_{\rm min}=\frac{\pi}{\sqrt{2}}\frac{GM_{\rm BH}}{\Delta E^{3/2}} \sim 40 \rm d 	\Bigl(\frac{\it R_{\rm *}}{1 \rm R_{\rm \odot}}\Bigr)^{3/2}\Bigl(\frac{\it M_{\rm BH}}{10^6 \rm M_{\rm \odot}}\Bigr)^{1/2}\Bigl(\frac{1\rm M_{\rm \odot}}{\it M_{\rm *}}\Bigr).             	\label{tmostbound}
\end{equation}
The returned debris then gradually circularise and form an accretion disc, and the subsequent fallback of the material onto the BH is thus driven by the viscous time. 

The rate of material returning at pericentre is $\dot M (t) =[(2\pi GM_{\rm BH})^{2/3}/3](dM/dE)t^{-5/3}$.
Here, $dM/dE$, which is the distribution of the bound debris per unit of specific orbital energy $E$ as a function of $E$ (i.e. $t$, the time since disruption) is not  exactly uniform (Lodato et al. 2009; Guillochon \& Ramirez-Ruiz 2013, 2015a), but, to a first approximation, we can consider it as such (Rees 1988; Evans \& Kochanek 1989). Considering that half of the stellar debris is bound to the BH after disruption and using Eq. \ref{tmostbound}, the returning rate trend starting from $t_{\rm peak}\sim 1.5 t_{\rm min}$ (Evans \& Kochanek 1989; Lodato et al. 2009) reads
\begin{multline}
\dot M (t) \sim \frac{(2\pi GM_{\rm BH})^{2/3}}{3}\frac{M_{\rm *}/2}{\Delta E} t^{-5/3}\sim \frac{2}{3} \frac{M_{\rm *}/2}{t_{\rm min}}\Bigl(\frac{\it t}{t_{\rm min}}\Bigr)^{-5/3}\sim \\
\sim 3.04 \rm M_{\rm \odot} \rm yr^{-1} \Bigl(\frac{\it R_{\rm *}}{1 \rm R_{\rm \odot}}\Bigr)\Bigl(\frac{\it M_{\rm BH}}{10^6 \rm M_{\rm \odot}}\Bigr)^{1/3} \Bigl(\frac{\it M_{\rm *}}{1 \rm M_{\rm \odot}}\Bigr)^{1/3}\Bigl(\frac{\it t}{0.109 \rm yr}\Bigr)^{\rm -5/3}.
\label{Mdotapprox}
\end{multline}
If the viscous time $t_{\rm \nu}\sim (2^{5/2} R_{\rm *}^{3/2})/(\sqrt{G}M_{\rm *}^{1/2}\alpha h^2)(r_{\rm t}/r_{\rm p})^{-3/2}$ (Li et al. 2002), where $\alpha$ is the viscous disc parameter and $h$ the disc half-height divided by its radius, is smaller than $t_{\rm min}$, meaning that if their ratio $t_{\rm \nu}/t_{\rm min}\sim 0.025 (\alpha/0.1)^{-1} h^{-2} (10^6 \rm M_{\rm \odot}/\it M_{\rm BH})^{\rm 1/2} (M_{\rm *}/ \rm1 M_{\rm \odot})^{1/2}(\it r_{\rm t}/r_{\rm p})^{\rm -3/2}$
is low, the rate of debris returning at pericentre $\dot M (t)$ coincides to a first approximation\footnote{For very deep encounters, relativistic effects may change the debris evolution from the one discussed here (e.g. Dai et al. 2015). For example, in-plane relativistic precession probably causes the stream of debris to self-cross, which speeds up its circularisation (Shiokawa et al. 2015; Bonnerot et al. 2016), while nodal relativistic precession deflects the debris out of their orbital plane, which delays self-intersection and circularisation (Guillochon \& Ramirez-Ruiz 2015b).} with the rate of material accreting onto the BH. The thicker the accretion disc (i.e. $h\sim 1$), the better this approximation. We also note that when the parameter $\beta=r_{\rm t}/r_{\rm p}$ approaches unity (it is $\gtrsim 1$ for total disruptions), this ratio is accordingly reduced. 

The luminosity produced by accretion can therefore be evaluated as
\begin{eqnarray}\nonumber
L (t) \sim \eta \dot M (t) c^2\sim 1.7\times 10^{46} \frac{\rm erg}{\rm s} \frac{\eta}{0.1}\Bigl(\frac{\it R_{\rm *}}{1 \rm R_{\rm \odot}}\Bigr) \Bigl(\frac{\it M_{\rm BH}}{10^6 \rm M_{\rm \odot}}\Bigr)^{1/3} \times \\ \times \Bigl(\frac{\it M_{\rm *}}{1 \rm M_{\rm \odot}}\Bigr)^{1/3} \Bigl(\frac{\it t}{0.109 \rm yr}\Bigr)^{\rm -5/3}
\end{eqnarray}
(where $\eta$ is the radiation efficiency), again starting from $t_{\rm peak}$. The peak luminosity $L_{\rm peak}$ turns out to be a factor of  $\sim 130$ super-Eddington for $M_{\rm BH}=10^6 \rm M_{\rm \odot}$, $M_{\rm *}=1 \rm M_{\rm \odot}$, $R_{\rm *}=1 \rm R_{\odot}$ and $\eta=0.1$, which means that probably a fraction of the accreted mass is ejected in a wind from the disc (Strubbe \& Quataert 2009; Lodato \& Rossi 2010). 
If we assume that $L_{\rm peak}$ roughly coincides with the Eddington luminosity and know the unabsorbed peak flux $F_{\rm peak}$, then we can evaluate the host galaxy distance $d$ as a function of $M_{\rm BH}$ as
\begin{equation}
d\sim 0.1 \rm Mpc \Bigl(\frac{\it M_{\rm BH}}{1 \rm M_{\rm \odot}}\Bigr)^{1/2}\Bigl(\frac{10^{-10}\rm erg/ \rm s/ \rm cm^{2}}{\it F_{\rm peak}}\Bigr)^{1/2}	\label{dist}.
\end{equation} 
In addition to using methods which require optical spectra (e.g. Nelson 2000; Marconi \& Hunt 2003; Sani et al. 2010), $M_{\rm BH}$ may then be approximately estimated from Eq. \ref{dist} if $d$ is known.

\section{XMMSL1J063045.9-603110} \label{known}
\begin{table*}
\centering
\caption{XMMSL1J063045 counterpart AB magnitudes measured by Kann et al. (2011) from GROND observations. \label{grond_table}}
\begin{tabular}{c | c c c c c c c }
Filter & g' & r' & i' & z' & J & H & K\\
& & & & & & & \\
\hline
& & & & & & & \\
m & $18.4\pm0.1$ & $19.5\pm0.1$ & $19.6\pm0.1$ & $20.1\pm 0.1$ & $20.3\pm0.2$ & $20.9\pm0.4$ & $>20$ \\
\end{tabular}
\end{table*}
\begin{figure*}
\centering
\caption{Left panel: Image from the DSS catalogue before the XMMSL1J063045 detection in X-rays. No counterparts are visible within the XMMSL1J063045 Slew error circle (green circle). Central panel: Swift UVOT uvw1 filtered image immediately subsequent to XMMSL1J063045 detected X-ray ignition. A possible counterpart appears within the XMMSL1J063045 Slew error circle (green) and within $5''$ from the source UVOT position (red circle). Right panel: Swift UVOT uvvv filtered image, obtained about four years after the uvw1 filtered observation. No counterparts are visible either within the XMMSL1J063045 Slew error circle (green circle) or within $5''$ from the source UVOT position (red circle). The images are $3.4' \times 3.4'$, north is up, east is left.
 \label{dss_image}}
\includegraphics[width=19.cm]{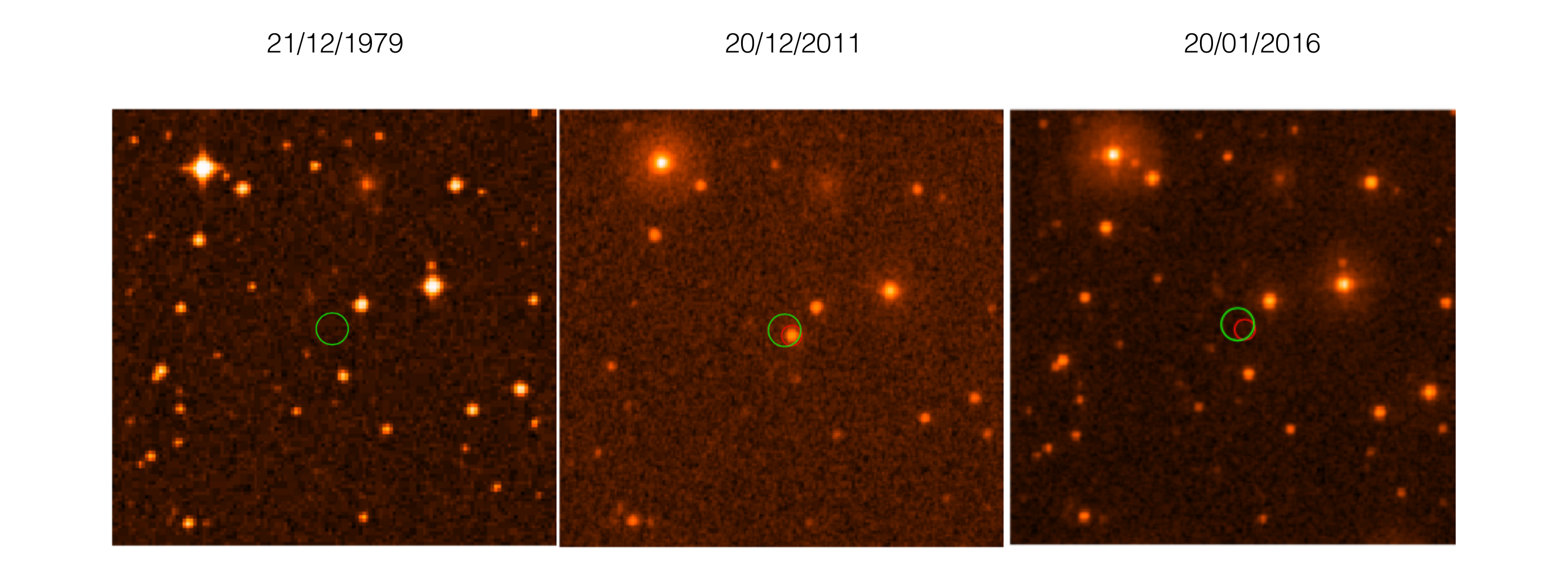} 
\end{figure*}
\begin{table*}
\centering
\caption{Log of all the XMMSL1J063045 X-ray observations. Pointed observations in italics have not previously been reported in the literature. The last Swift XRT observation in bold italics was required to check the current state of the source. XMM-Newton Slew (SL) count rates (0.2-2keV) were adopted from Read et al. (2011a), Swift count rates (0.3-2keV) from the online catalogue at the webpage http://www.swift.ac.uk/1SXPS/1SXPS\%20J063045.2-603110, except for the last observation, whose upper limit on the count rate was evaluated using the \texttt{XIMAGE} task \texttt{sosta}. The XMM-Newton pointed observed count rate (0.2-2keV) was extracted from the corresponding source spectrum. The 0.2-2keV unabsorbed fluxes are estimated as reported in Sect. \ref{variability}. \label{observations_table}}
\begin{tabular}{c | c | c | c | c | c | c}
Instrument & Obs. ID & Start time & Start time & Exp. time & Source count rate &  0.2-2keV Unabs. flux\\
& & dd/mm/yy & MJD & s & ct $\rm s^{-1}$ & erg $\rm s^{-1}$ $\rm cm^{-2}$\\
& & & & & & \\
\hline
& & & & & & \\
\small{XMM-Newton (SL) EPIC-PN} & & 14/08/2002 & 52500 & & $< 0.52$ & $< (2.8\pm 0.3)\times 10^{-12}$ \\
\small{XMM-Newton (SL) EPIC-PN} & & 18/11/2008 & 54788 & & $< 1.76$ & $< (9.5\pm 0.9)\times 10^{-12}$  \\
\small{XMM-Newton (SL) EPIC-PN} & & 01/12/2011 & 55896 &  & 32.6 & $(1.7\pm 0.2)\times 10^{-10}$ \\
\small{Swift XRT} & 00032225001 & 20/12/2011 & 55915 & $2.9\times 10^3$ & $(6.4\pm0.5)\times 10^{-2}$ & $(1.2\pm 0.2)\times 10^{-11}$ \\
\it{\small{XMM-Newton EPIC-PN}} & \it{0679381201} & \it{22/12/2011} & \it{55917} & $\it{6.5\times 10^3}$ & $\it{1.94\pm0.02}$ & $\it (8.7\pm 0.9)\times 10^{-12}$ \\
\it{\small{Swift XRT}} & \it{00032225002} & \it{06/01/2012} &  \it 55932 & \it{471} & $\it{2.5^{+0.9}_{-0.7}\times 10^{-2}}$ &  $\it (4.6^{+1.9}_{-1.5})\times 10^{-12}$ \\
\it{\small{Swift XRT}} & \it{00032225003} & \it{11/01/2012} & \it 55937 & \it{579} & $\it{1.6^{+0.7}_{-0.5}\times 10^{-2}}$ &  $\it (2.9^{+1.4}_{-1.0})\times 10^{-12}$  \\
\it{\small{Swift XRT}} & \it{00032225004} & \it{12/01/2012} & \it 55938 & $\it{2.1\times 10^3}$ & $\it{8.0^{+2.4}_{-2.0}\times 10^{-3}}$ &  $\it (1.5^{+0.5}_{-0.4})\times 10^{-12}$\\
\it \textbf{\small{Swift XRT}} & \it \textbf{00032225005} & \it \textbf{20/01/2016} & \it \textbf{55917} & \boldmath $\it 4.3 \times 10^3$ & \boldmath $\it < 2.4\times 10^{-3}$ & \boldmath $\it < (4.4^{+0.8}_{-0.7})\times 10^{-13}$\\
\end{tabular}
\end{table*}

On December 1, 2011, the new point-like source XMMSL1J063045.9-603110 (hereafter XMMSL1J063045) was detected to be bright in the X-ray sky probed by the XMM-Newton Slew Survey (Saxton et al. 2008), at RA=06:30:45.9, DEC=-60:31:10 ($8''$ error circle, 1$\sigma$ confidence level). The source was soft, with essentially no emission above 2keV. Fitting the Slew X-ray spectrum, Read et al. (2011a) estimated an absorption of $N_{\rm H}=0.11\times 10^{22} \rm cm^{-2}$ ($\sim$ 2.1$N_{\rm H_{\rm Gal}}$), a blackbody temperature of $T_{\rm bb}=85 \rm eV$, and an absorbed 0.2-2keV EPIC-PN flux of $4.0\times 10^{-11} \rm erg$ $\rm s^{-1}$ $\rm cm^{-2}$, starting from a count rate of $32.6 \rm ct$ $\rm s^{-1}$. When the same spectral model is assumed, the upper limits obtained from two previous XMM-Newton slews over this position, which are $< 0.52 \rm ct$ $\rm s^{-1}$ (14/08/2002) and $< 1.76 \rm ct$ $\rm s^{-1}$ (18/11/2008), give absorbed 0.2-2keV EPIC-PN fluxes of $< 6.4\times 10^{-13} \rm erg$ $\rm s^{-1}$ $\rm cm^{-2}$ and $< 2.2 \times 10^{-12} \rm erg$ $\rm s^{-1}$ $\rm cm^{-2}$. These are factors of more than 63 and 18 below the bright Slew detection, respectively. This flux gap, together with the non-detection of previous lower energy counterparts (Fig. \ref{dss_image}, left panel), 
led Read et al. to suggest that XMMSL1J063045 might be a new nova. 

On December 18, 2011, Kann et al. (2011) identified an object at RA=06:30:45.45, DEC=-60:31:12.8 with an error of $\pm 0.3''$, which is fully within the XMMSL1J063045 Slew error circle, based on simultaneous filtered observations of the XMMSL1J063045 field with the optical telescope GROND. The authors suggested it might be the counterpart of XMMSL1J063045. Table \ref{grond_table} summarises the AB magnitudes they measured with different filters. 
The faint brightness (the Galactic reddening at this position is only E(B-V)=0.07; Schlegel et al. 1998), coupled with the very blue g'-r' colour evaluated by Kann et al., is atypical for a nova, which discards the classification suggested by Read et al. (2011a) and favours an accretion disc hypothesis. Fitting the source spectrum with a -2 power law, Kann et al. found a deviation in the g' band ($4000-5400 \rm \AA$), which they interpreted as due to a strong HeII emission ($\lambda_{\rm HeII}=4685 \rm \AA$). 

On December 20, 2011, the Swift satellite also revealed a soft X-ray source coincident with the object detected with GROND. To be specific, the Swift/UVOT UVW1 ($2000-3300 \rm \AA$) source position is RA=06:30:45.42, DEC=-60:31:12.54 ($0.44''$ error circle, $90\%$ confidence level). From fitting the XRT spectrum with $N_{\rm H}\equiv \it N_{\rm H_{\rm Gal}}=\rm 5.11\times 10^{20} \rm cm^{-2}$, Read et al. (2011b) found a blackbody temperature of $T_{\rm bb}=48\pm5 \rm eV$ and an absorbed 0.2-2keV flux of $3.4^{+0.8}_{-1.2}\times 10^{-12} \rm erg$ $\rm s^{-1}$ $\rm cm^{-2}$, which is a factor of about 12 below the XMM-Newton Slew bright flux. 

Despite the peculiar features of this source (soft X-ray thermal spectrum, blackbody temperature decrease, high and rapid X-ray flux decay, accretion-disc-like optical-UV spectrum), nothing else can be found in the literature. 

Table \ref{observations_table} lists the whole of the XMMSL1J063045 X-ray observations, also including four observations that were not previously reported in the literature (in italics) and another that was specifically required to check the current state of the source (in bold italics). 
In the following sections we present our X-ray data analysis and discuss the possible nature of the source. 

\section{X-ray data reduction} \label{data_reduction}
\subsection{XMM-Newton Slew Survey}
XMMSL observations were carried out with all the three imaging EPIC cameras (PN, MOS1, and MOS2) onboard XMM-Newton, but the high Slew speed and the slow readout time of MOS1 and MOS2 (Turner et al. 2001) prevent MOS data from being analysed. Therefore Read et al. (2011a) analysed only EPIC-PN data (Str$\rm \ddot u$der et al. 2001) of XMMSL1J063045. XMMSL data are very difficult to analyse, and for this reason we rely on the analysis of Read et al. (Sect. \ref{known}). 

\subsection{Swift}
The composite XRT spectrum of XMMSL1J063045, obtained by grouping the four Swift XRT observations close in time listed in Table \ref{observations_table}, can be directly downloaded from the online Swift source catalogue (Evans et al. 2014).\footnote{http://www.swift.ac.uk/1SXPS/spec.php?sourceID=1SXPS$\textunderscore$\\J063045.2-603110} No emission above 2keV is observed. Hence, the source count rates reported in Table \ref{observations_table} for each XRT observation\footnote{http://www.swift.ac.uk/1SXPS/1SXPS\%20J063045.2-603110} can be approximately associated with the 0.3-2keV energy band. 
We binned spectral data with the \texttt{grppha} tool of \texttt{HEASoft} (v.6.17) to a minimum of one photon per channel of energy, given the low number of photons, and we adopted Cash-statistics when fitting data. 

The last XRT observation reported in bold italics in Table \ref{observations_table} was reprocessed using \texttt{xrtpipeline} (v.0.13.2), and its corresponding upper limit on the source count rate was evaluated using the \texttt{XIMAGE} (v.4.5.1) task \texttt{sosta}.

\subsection{XMM-Newton} \label{XMM_reduction}
The XMM-Newton pointed observation of XMMSL1J063045 (Table \ref{observations_table}) was carried out with all the three XMM-Newton EPIC cameras using the thin filter. We reprocessed data using \texttt{SAS} (v. 13.5.0). We filtered them for periods of high background flaring activity, setting the maximum threshold on the source light curve count rates at 0.4 (0.35)ct $\rm s^{-1}$ for the PN (MOS) camera. Data were also filtered with the FLAG==0 option, and only events with pattern $\leq$4 ($\leq$12) were retained. For all the three cameras, we extracted the source$+$background spectrum from a circular region of radius $40''$, centred on the source. We cleaned these spectra of the background, extracted from a circular region of radius $40''$ on the same CCD, free of sources and bad columns. \texttt{RMF} and \texttt{ARF} files were produced using the appropriate tasks. We binned the obtained source spectra to a minimum of 20 photons per channel of energy. Data were accumulated in the 0.2-2keV (0.3-2keV) energy band for the PN (MOS) camera. 

\section{X-ray spectral analysis} \label{data_analysis}
\begin{figure*}
\centering
\caption{PN (black), MOS1 (red) and MOS2 (green) soft X-ray spectra obtained from the XMM-Newton pointed observation of XMMSL1J063045 reported in Table \ref{observations_table}, fitted with an absorbed thermal accretion disc model, in agreement with Read et al. (2011a) and Kann et al. (2011; Sect. \ref{known}). Residuals in terms of $\Delta \chi$ are plotted in the lower panel with corresponding colours and are well distributed throughout the concerned energy band. 
 \label{xmm_fit}}
\includegraphics[width=15.cm, angle=0]{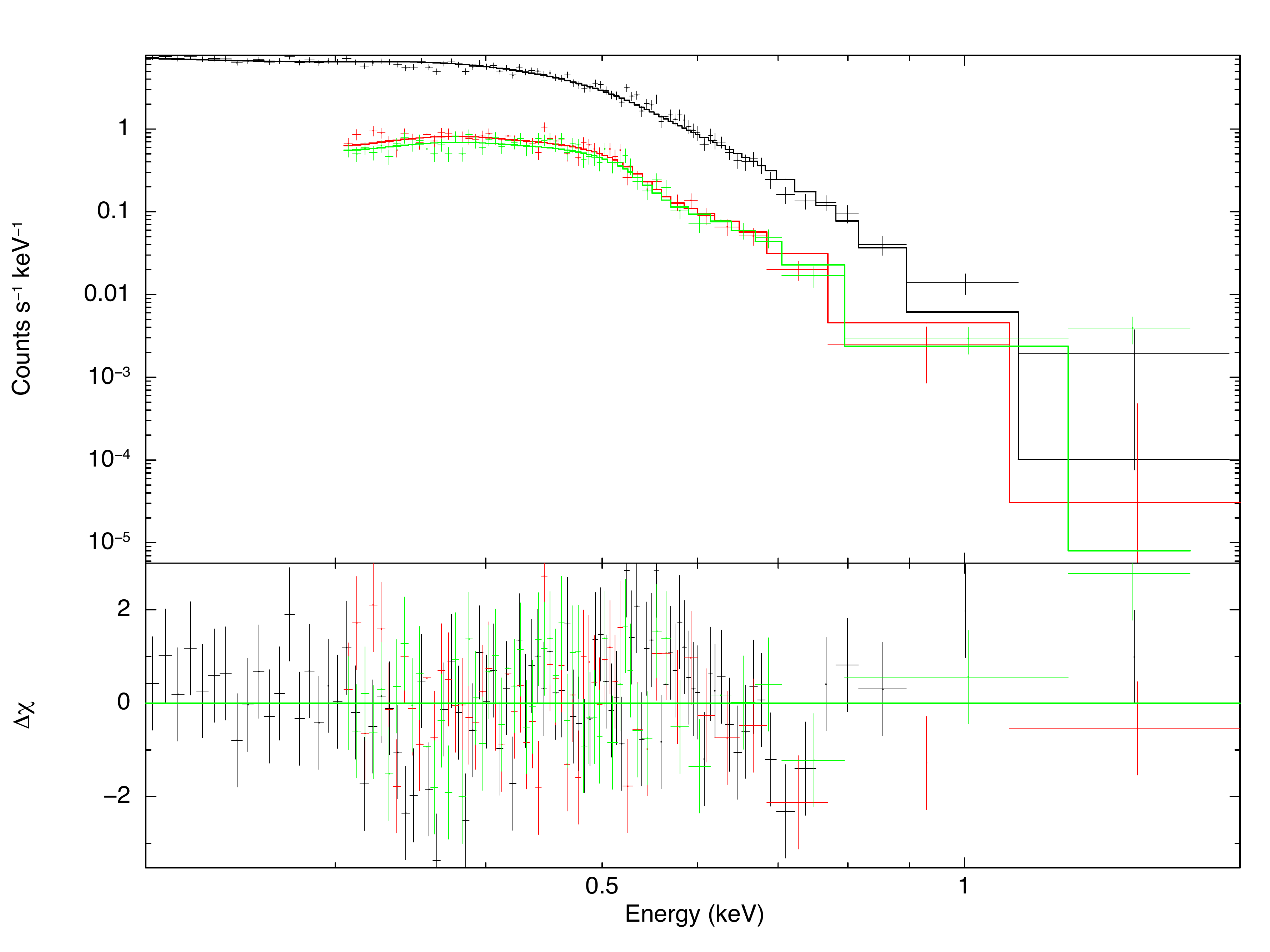}
\end{figure*}

From the XMM-Newton pointed observation of XMMSL1J063045 reported in Table \ref{observations_table} we obtained three distinct soft X-ray spectra, one for each EPIC camera (Sect. \ref{XMM_reduction}).\footnote{We also inspected the XMM-Newton RGS data, but failed to find emission or absorption lines.} We fit them together with an absorbed (using \texttt{TBABS}) power-law model from the package \texttt{XSPEC} (v.12.9.0), tying together all the column densities and all the photon indexes. The photon index is extremely high, with $\Gamma=9.8\pm 0.2$ (all errors are determined at the $90\%$ confidence level), and the column density $N_{\rm H}=(17.41\pm 0.31)\times 10^{20} \rm cm^{-2}$ significantly exceeds the Galactic value $N_{\rm H_{\rm Gal}}=5.11\times 10^{20} \rm cm^{-2}$. We obtained a  $\chi^{2}$-statistics value of 359.7 with 200 degrees of freedom (dof) and a corresponding null hypothesis probability (nhp) of $10^{-11}$. 
An absorbed bremsstrahlung model provides significantly better results: $\chi^2=248.7$ with 200 dof and a corresponding nhp of $1.1\%$. Even better results are obtained with an absorbed thermal accretion disc (\texttt{diskbb}) model, which also unifies the thermal nature of the XMMSL1J063045 X-ray emission, as identified by Read et al. (2011a), and the accretion disc appearance inferred by Kann et al. (2011) from the source optical emission (Sect. \ref{known}). This model, with column densities $N_{\rm H}$ and \texttt{diskbb} temperatures $T$ tied together, gives $N_{\rm H}=7.79^{+0.55}_{-0.53}\times10^{20} \rm cm^{-2}$, somewhat in excess of the Galactic value, and $T=59\pm1 \rm eV$, returning a $\rm \chi^{2}$-statistics value of 237.5 with 200 dof and a corresponding nhp of $3.7\%$ (Fig. \ref{xmm_fit}). 

Given the low number of photons that appear in the Swift X-ray observations of XMMSL1J063045, we again fit the composite XRT spectrum of the source with the \texttt{diskbb} model, fixing the column density to $N_{\rm H}=7.79\times10^{20} \rm cm^{-2}$. This fit gives $T=58^{+6}_{-5} \rm eV$, returning a $\chi^{2}$-statistics value of 20.86 with 37 dof, assessed using the Churazov-weighted $\chi^{2}$-statistics (Churazov et al. 1996) applied to the best fit with Cash-statistics.

\section{X-ray flux variability} \label{variability}
We now modelled the XMMSL1J063045 (soft) X-ray emission. To do this, we converted the count rates associated with the source X-ray observations into unabsorbed fluxes (Table \ref{observations_table}). 

From the XMMSL observations, we considered the 0.2-2keV absorbed fluxes that result from the analysis of Read et al. (2011a; Sect. \ref{known}). The conversion factor aimed at obtaining the corresponding 0.2-2keV unabsorbed fluxes can be easily estimated based on the Read et al. best spectral fit of the more recent XMMSL data, setting $N_{\rm H}=0$. By applying this correction factor to all the three XMMSL observations, we obtained the 0.2-2keV unabsorbed fluxes reported in Table \ref{observations_table}. The assumed relative uncertainty on fluxes for XMMSL observations is $10\%$. 

The 0.2-2keV unabsorbed fluxes associated with Swift XRT count rates were computed by means of the conversion factor $(1.83^{+0.31}_{-0.27})\times 10^{-10} \rm erg$ $\rm s^{-1}$ $\rm cm^{-2}$ ($\rm ct$ $\rm s^{-1}$)$^{-1}$ extracted from the unabsorbed thermal accretion disc spectral model applied to the composite XRT spectrum. In particular, we applied the conversion factor obtained by summing four XRT observations to each XRT observation, assuming the same spectral model also for the last observation in Table \ref{observations_table}. Uncertainties on the unabsorbed fluxes result from error propagation. We used the same method to compute the 0.2-2keV (EPIC-PN) unabsorbed flux corresponding to the source XMM-Newton pointed observation and its uncertainty. 
A relative systematic uncertainty of $10\%$, as for XMMSL fluxes, was also considered for all observations according to error propagation, thus justifying the comparison of data carried out with different satellites. 

\begin{figure*}
\centering
\caption{XMMSL1J063045 X-ray flux light curve fitted with a $-5/3$ power law (red solid line; $\chi^2_{\rm red}=2.5$, $\rm nhp=4.2\%$), as for tidal disruption events, and with a free power law (blue dashed line; $\chi^2_{\rm red}=3.2$, $\rm nhp=2.2\%$), which gives a decay index of $-1.71\pm 0.4$, fully in agreement  with -5/3. The right top panel also includes the upper limits on flux.   \label{lc_figure}}
\includegraphics[width=15.cm, angle=0]{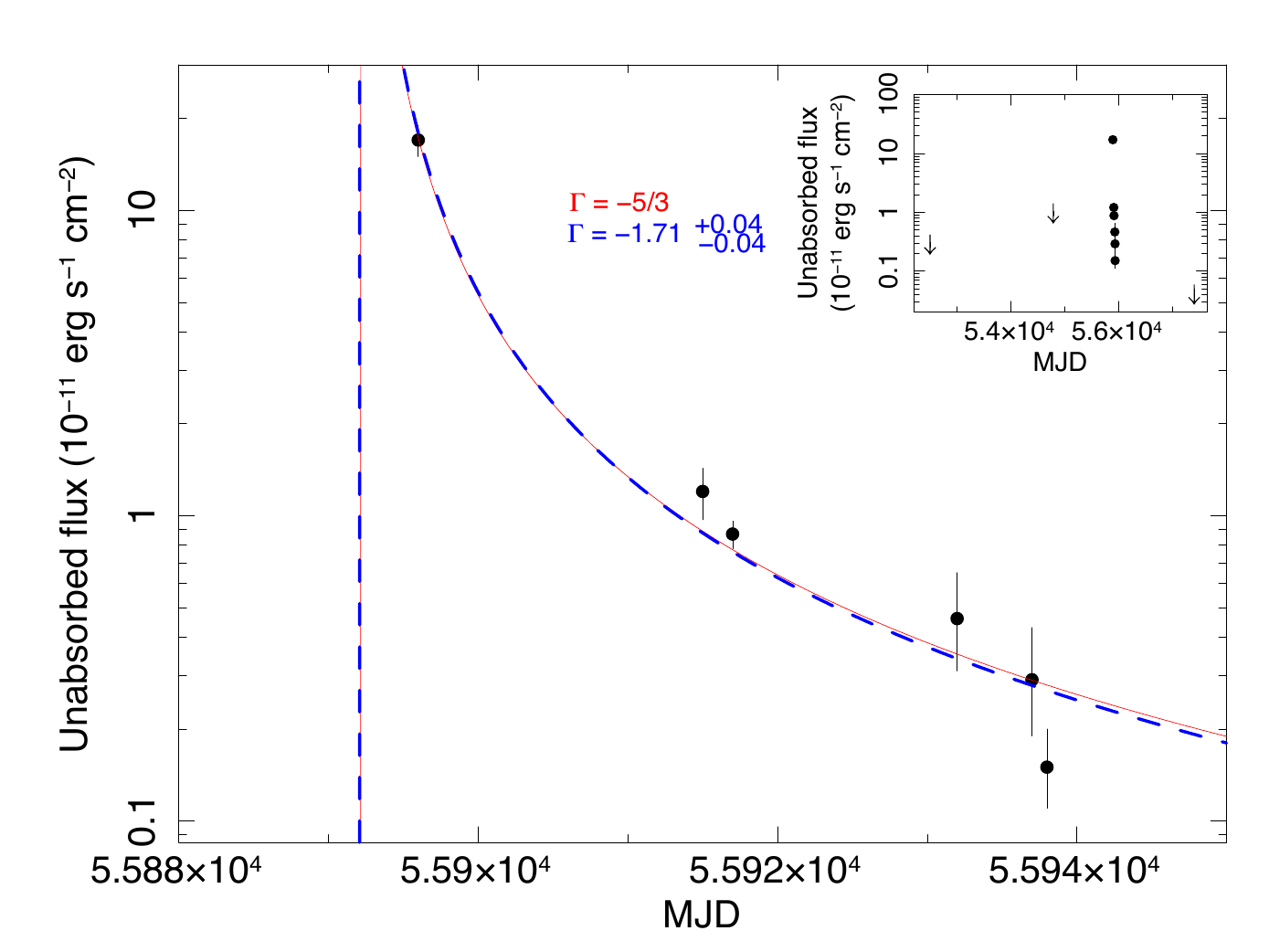} \\
\end{figure*}

Figure \ref{lc_figure} shows the XMMSL1J063045 X-ray flux light curve without the XMMSL and XRT upper limits. The right top panel shows the light curve with these limits. We fit the unabsorbed fluxes with a $(t_{\rm MJD}-t_{\rm 0})^{-5/3}$ power law, typical of a tidal disruption event (Sect. \ref{fundamentals}), with $t_{\rm 0}$ being a characteristic parameter (red solid line). The obtained $\chi^2$ is 9.9 with 4 dof ($\chi^2_{\rm red}=2.5$) and the corresponding nhp is $4.2\%$. A fit with a free power-law index (blue dashed line) gives $\chi^2=9.6$  with 3 dof ($\chi^2_{\rm red}=3.2$) and a corresponding nhp of $2.2\%$. Moreover, the power-law index is $-1.71\pm 0.04$, fully in agreement with $-5/3$. The XRT upper limit is lower than the last fitted flux value, which means that the source is still quiescent in the X-ray band today. 

In addition to the XMMSL1J063045 X-ray light curve, we found a downward trend in the temperatures derived from spectral analysis. In particular, we simulated the XMMSL source spectrum analysed by Read et al. (Sect. \ref{known}; 01/12/2011) using the \texttt{fakeit} option from the package \texttt{XSPEC} and grouped it to a minimum of 20 photons per channel of energy. Fitting it with an absorbed thermal accretion disc model (\texttt{TBABS}*\texttt{diskbb} from \texttt{XSPEC}), we obtained a \texttt{diskbb} temperature of 97eV. The following Swift composite spectrum (Dec. 2011/Jan. 2012) and XMM-Newton pointed observation (22/12/2011) show \texttt{diskbb} lower temperatures of $58^{+6}_{-5} \rm eV$ and $59\pm 1 \rm eV$, respectively. 

The soft X-ray thermal accretion disc emission of the source together with its temperature decrease and its high and rapid $t^{-5/3}$ flux decay (a factor of about 115 in only a month and a half) are all evidence of the probable TD nature of XMMSL1J063045.

\section{Swift UVOT data} \label{UVOT}
\begin{table*}
\centering
\caption{XMMSL1J063045 counterpart AB magnitudes assessed from Swift UVOT filtered observations carried out after the source ignition in X-rays. No corrections for Galactic extinction are applied. \label{UVOT_magnitudes_table}}
\begin{tabular}{c | c c c c c }
Start time (dd/mm/yy) & 20/12/2011 & 06/01/2012 & 11/01/2012 & 12/01/2012 & 20/01/2016 \\
& & & & & \\
\hline
& & & & & \\
Filter & uvw1 & uvuu & uvw2 & uvm2 & uvvv\\
& & & & & \\
\hline
& & & & & \\
m & $18.77\pm0.03$ & $18.21\pm0.04$ & $19.17\pm0.06$ & $19.31\pm 0.05$ & $> 20.07$ \\
\end{tabular}
\end{table*}
A further comment on XMMSL1J063045 concerns its activity at lower energies. The left panel of Fig. \ref{dss_image} shows no lower energy counterparts of the source before it lights up in X-rays. Swift XRT observations reported in Table \ref{observations_table}, subsequent to the source X-ray ignition, are all coupled with Swift UVOT observations, each one carried out using only one filter (uvw1, uvuu, uvw2, uvm2, and uvvv). The central panel of Fig. \ref{dss_image} shows the XMMSL1J063045 uvw1 filtered field and the lighting up of a probably lower energy counterpart of the source, immediately after its X-ray activity. The right panel of Fig. \ref{dss_image} shows no lower energy counterparts of the source in its uvvv filtered field, about four years after the source detection in X-rays. 
Table \ref{UVOT_magnitudes_table} also collects the source counterpart AB magnitudes associated with the five differently filtered UVOT observations, assessed with the \texttt{uvotsource} tool of \texttt{HEASoft} without correcting for Galactic extinction. 
This clearly is a soft X-ray/ optical-UV transient. 

\section{XMMSL1J063045 host galaxy} \label{cTD}
The main factor in stellar tidal disruptions is the destroyer BH, whose mass can be approximately related to the source luminosity distance $d$ through Eq. \ref{dist}. For XMMSL1J063045, the unabsorbed peak bolometric flux $F_{\rm peak}$ that appears in Eq. \ref{dist} can be inferred by fitting our simulated XMMSL spectrum  (Sect. \ref{variability}) with the best spectral model of Read et al. (2011a) by setting $N_{\rm H}=0$ and extrapolating data in the 0.01-10keV energy band. The flux is $2.8\times 10^{-10} \rm erg$ $\rm s^{-1}$ $\rm cm^{-2}$, so that
\begin{equation}
d\sim 0.06 \rm Mpc \Bigl(\frac{\it M_{\rm BH}}{1 \rm M_{\rm \odot}}\Bigr)^{1/2}, \label{dMBH}
\end{equation}
which is 60Mpc (redshift $z$=0.014; $H_{\rm 0}=69.6 \rm km$ $\rm s^{-1}$ $\rm Mpc^{-1}$, $\Omega_{\rm m}=0.286$) for a BH of mass $10^6 \rm M_{\rm \odot}$. 
An upper limit on XMMSL1J063045 distance ($z$) can be assessed by imposing a maximum value for $M_{\rm BH}$ of $10^8 \rm M_{\rm \odot}$, as for tidally disrupted solar-type stars. This limit is $600\rm Mpc$ ($z$=0.13). We here assumed that the observed peak luminosity of the source coincides with its Eddington limit. On one hand, we are aware that this limit can be exceeded by a factor of several, but on the other hand, we note that the actual outburst peak is probably brighter than the value we infer from the bright Slew detection. Hence we consider the Eddington limit as an acceptable compromise between these two competing instances.   

\begin{figure*}
\centering
\caption{Zoom of a reduced 300 s ESO-NTT EFOSC2 recent image of the XMMSL1J063045 field in the V filter (09/01/2016, about four years after XMMSL1J063045 X-ray detection). The image is $0.9'  \times 0.9'$, north is up, east is left. A dim extended source, possibly the XMMSL1J063045 host galaxy, is visible within $2.5''$ from XMMSL1J063045 UVOT position (green circle). \label{NTT}}
\includegraphics[width=12.cm, angle=0]{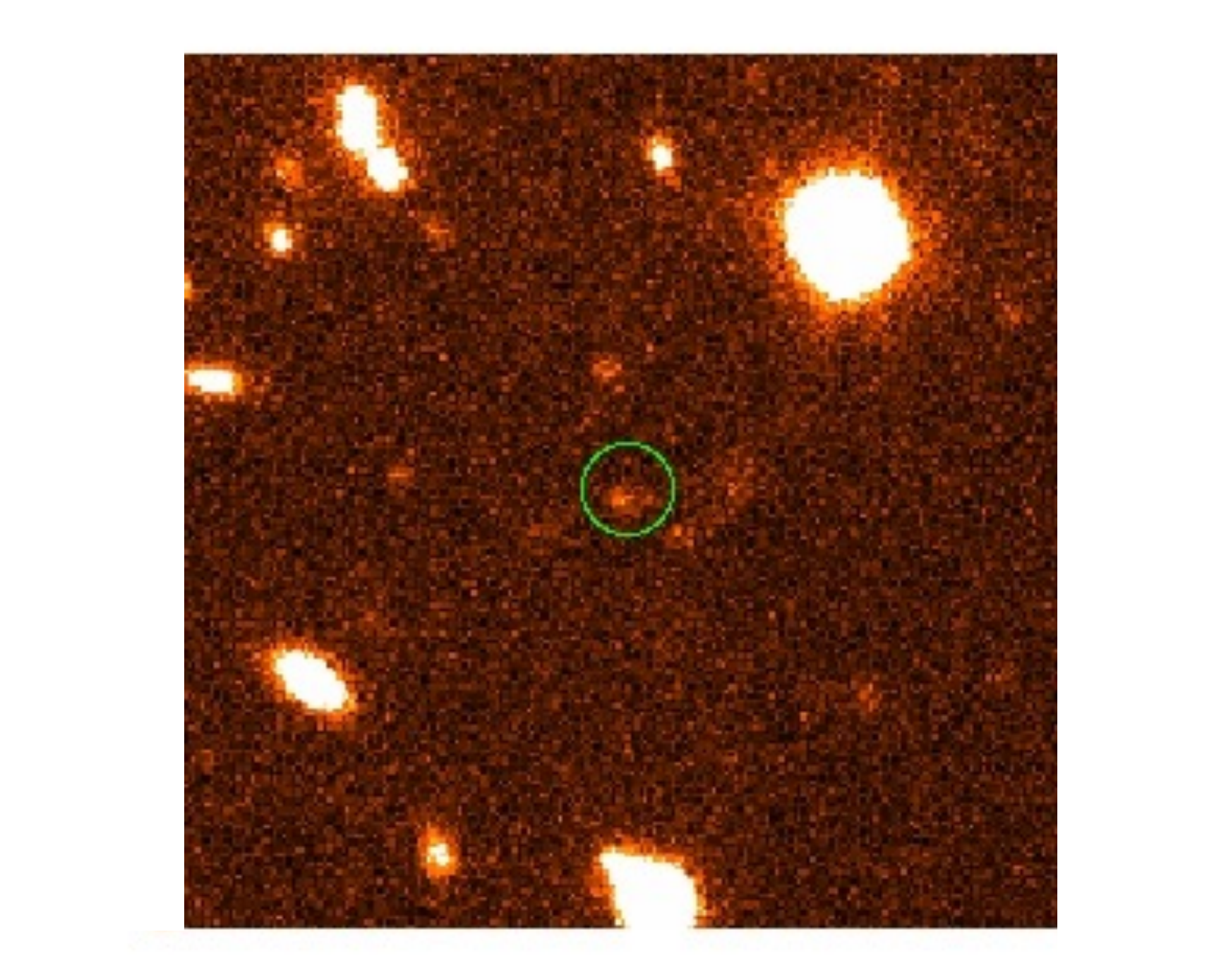} \\
\end{figure*}

The XMMSL1J063045 redshift and, consequently, luminosity distance might be inferred from its host galaxy spectroscopy, provided that there is a host galaxy, which should be in the case of TDs. No signs of it can be found in the DSS image (Fig. \ref{dss_image}, left panel) or in the Swift UVOT uvvv filtered image (Fig. \ref{dss_image}, right panel). 
On January 9, 2016, we obtained a deep 300 s V-band ESO-NTT image of the field of XMMSL1J063045. This was carried out with the EFOSC2 instrument. The observations were taken as part of the Public ESO Spectroscopic Survey of Transient Objects
(PESSTO\footnote{www.pessto.org}), and details of data products and reductions can be found in Smartt et al. (2015). Calibrating it  through the identification of four objects in the UVOT uvvv filtered image using the \texttt{uvotsource} tool of \texttt{HEASoft} and the \texttt{GAIA} software (v. 4.4.1), we find an object of apparent V magnitude $m_{\rm V}\sim 23.26\pm 0.27$ at the UVOT position of XMMSL1J063045 (Fig. \ref{NTT}, green circle). This probably is the XMMSL1J063045 (dim) host galaxy.
Currently, the dimmest (and smallest) galaxy associated with a TD is WINGS J134849.88+263557.7 in Abell1795 (Maksym et al. 2013; Maksym et al. 2014a; Donato et al. 2014). It is a faint ($m_{\rm V}=22.46$) dwarf ($r\sim 300 \rm pc$) galaxy lying at the same redshift of the cluster ($z=0.062$; $M_{\rm V}\sim-14.8$), possibly hosting an IMBH ($M_{\rm BH}=(2-5)\times 10^5 \rm M_{\rm \odot}$). The problem for XMMSL1J063045 is that no spectra deep enough of its dim host galaxy are currently available in which absorption lines might be identified to evaluate $z$. 

Using Eq. \ref{dMBH} to estimate the source luminosity distance $d_{\rm pc}$ (in parsec) from a BH of mass  $10^4-10^6 \rm M_{\rm \odot}$ and considering the relation between the absolute magnitude ($M_{\rm V}$) and the apparent magnitude ($m_{\rm V}$),
the dim host ends up with $M_{\rm V}\sim-5.7\div -10.7$. When $L_{\rm peak}$ is assumed to be ten times the Eddington luminosity (which is very high), $M_{\rm V}$ lies in the range $-8.2\div -13.2$. This value is at the level of the faintest dwarf spheroidal galaxies in our Milky Way (Sculptor has $M_{\rm V}=-10.7$) or of the brightest globular clusters (NGC5139 has $M_{\rm V}=-10.2$), opening the possibility of observing the first TD in a globular cluster and suggesting that IMBHs are present in the cores of at least some of them.  

\section{Summary and conclusions} \label{conclusions}
In a galactic nucleus, a star that approaches to the central black hole too closely may be tidally disrupted before it is fully absorbed into the black hole horizon. A fraction of the produced stellar debris is bound to the compact object, accreting onto it through an accretion disc and lighting it up through a characteristic flare (Rees 1988; Phinney 1989). 
Up to now, a limited number ($\sim 65$) of tidal disruption candidates have been observationally identified,\footnote{https://astrocrash.net/resources/tde-catalogue/} mainly in the optical-UV and soft X-ray bands (e.g. Gezari 2012 and references therein; Komossa 2015 and references therein), both involving massive galaxies, that is, supermassive black holes (Kormendy \& Richstone 1995; Kormendy \& Gebhardt 2001), and dwarf galaxies, which possibly host intermediate-mass black holes at their centres (Ghosh et al. 2006; Maksym et al. 2013; Maksym et al. 2014a; Donato et al. 2014; Maksym et al. 2014b). The discovery of new tidal disruption candidates would certainly improve our understanding of the physics behind them, and if they were detected in dwarf galaxies, we might be able to determine plausible destroyer intermediate-mass black holes. This class of black holes is currently under study (e.g. Ptak \& Griffiths 1999; Davis \& Mushotzky 2004; Wolter et al. 2006; Greene \& Ho 2007; Farrell et al. 2009; Irwin et al. 2010; Jonker et al. 2010; Krolik \& Piran 2011; Jonker et al. 2013) as the connecting bridge between stellar-mass and supermassive black holes and the raw material that makes up supermassive black holes (e.g. Volonteri 2010). 

On December 1, 2011 the new point-like source XMMSL1J063045.9-603110 was detected to be bright in the soft X-rays, with an underlying thermal emission (Read et al. 2011a). An accretion-disc nature was suggested  (Kann et al. 2011). After about twenty days, XMMSL1J063045.9-603110 was also observed by the Swift satellite, again producing a soft X-ray thermal emission, a factor of about 10 below its first detection (Read et al. 2011b). We reported here a comprehensive data analysis of all the available 
X-ray (Table \ref{observations_table}) and lower energy data (Sect. \ref{UVOT}) of XMMSL1J063045.9-603110. We suggest that the source is a tidal disruption event. It showed an accretion-disc-like thermal spectrum in the soft X-rays (Fig. \ref{xmm_fit}) together with a high and rapid flux decay (a factor of $\sim 115$ in only a month and a half) that is well modelled by a power law of index -5/3 (Fig. \ref{lc_figure}). Moreover, the source also blazed up at lower energies (Fig. \ref{dss_image}), even if it slightly lags behind the X-ray flaring. 

We reject the hypothesis that XMMSL1J063045.9-603110 is a Galactic nova (Galactic latitude $b=-26$). The softness of the source X-ray spectrum would require such a nova to be in a super-soft state (SSS). To reach this state, the source would need to radiate at Eddington ($\sim 1.3\times 10^{38} \rm erg$ $\rm s^{-1}$) and would have to lie at $d \sim 65 \rm kpc$.  
Furthermore, the source magnitude variation is about 5 mag, which is too small for a typical nova, and its quiescent magnitude (Sect. \ref{cTD}) is too high for a typical super-soft nova (the dimmest one, GQ Mus, has a quiescent V magnitude of $\sim 18$; e.g. Warner 2002). The reported magnitudes are not enhanced by Galactic extinction, given that this is very low at the source position (Schlegel et al. 1998). 
 We also discard the idea that XMMSL1J063045.9-603110 is an AGN because it would have been detected in all the observations if that were the case (see also the discussion in Campana et al. 2015). 
  
Based on the hypothesis that XMMSL1J063045.9-603110 is a candidate tidal disruption event, the low \texttt{diskbb} temperature that characterises the source ($\sim 100 \rm eV$; Sect. \ref{variability}) would call for a $\sim 10^4 \rm M_{\rm \odot}$ destroyer black hole, assuming that it accretes at the Eddington rate, or a $\sim 10^5 \rm M_{\rm \odot}$ black hole, assuming that it accretes at ten times the Eddington rate. It might be a tidal disruption event in a very dim dwarf galaxy of even in a very bright globular cluster ($M_{\rm V}\sim-10$), which then could host a black hole at their centre. Globular clusters typically do not wander alone in the cosmos, but are associated with a parent galaxy. Figure \ref{NTT} shows that the field of XMMSL1J063045.9-603110 is sparsely crowded, but there is something around it, possibly also a parent galaxy. Spectroscopic observations of XMMSL1J063045.9-603110 host will provide a clearer answer.

\begin{acknowledgements}
This work is based (in part) on observations collected at the European Organisation for Astronomical Research in the Southern Hemisphere, Chile as part of PESSTO (the Public ESO Spectroscopic Survey of Transient Objects) ESO program 188.D-3003, 191.D-0935. We thank P. D'Avanzo and S. J. Smartt for useful comments and discussions. We acknowledge M. C. Baglio for useful comments about image calibration. 
We also thank the anonymous referee for valuable comments on the manuscript and constructive suggestions.
\end{acknowledgements}


\begin{thebibliography}{}
\makeatletter
\def\@biblabel#1{}
\let\old@bibitem\bibitem

\bibitem{Alexander2012}
Alexander, T. 2012, in European Physical Journal Web of Conferences, p. 5001 (arxiv:1210.0582), doi: 10.1051/ep jconf/20123905001

\bibitem{Begelman2006}
Begelman, M. C., Volonteri, M. \& Rees, M. J. 2006, MNRAS, 370, 298

\bibitem{Bloom2011}
Bloom, J. S., Giannios, D. et al. 2011, Science, 333, 203

\bibitem{Bonnerot2016}
Bonnerot, C., Rossi, E. M., Lodato, G. \& Price, D. J. 2016, MNRAS, 455, 2253

\bibitem{Burrows2011}
Burrows, D. N., Kennea, J. A. et al. 2011, Nature, 476, 421

\bibitem{Campana2015}
Campana, S., Mainetti, D. et al. 2015, A\&A, 581, A17

\bibitem{Cenko2012}
Cenko,  S. B., Krimm, B. et al. 2012, ApJ, 753, 77

\bibitem{Churazov1996}
Churazov, E., Gilfanov, M. et al. 1996, ApJ, 471, 673

\bibitem{Dai2015}
Dai, L., McKinney, J. C. \& Miller, M. C. 2015, ApJ, 812, L39

\bibitem{Davis2004}
Davis, D. S. \& Mushotzky, R. F. 2004, ApJ, 604, 653

\bibitem{Donato2014}
Donato, D., Cenko, S. B. et al. 2014, ApJ, 781, 59

\bibitem{Donley2002}
Donley, J. L., Brandt, W. N., Eracleous, M. \& Boller, Th. 2002, AJ, 124, 1308

\bibitem{Evans1989}
Evans, C. R. \& Kochanek, C. S. 1989, ApJ, 346, L13

\bibitem{Evans2014}
Evans, P. A., Osborne, J. P. et al. 2014, ApJS, 210, 8

\bibitem{Farrell2009}
Farrell, S. A., Webb, N. A. et al. 2009, Nature, 460, 73

\bibitem{Frank1976}
Frank, J. \& Rees, M. J. 1976, MNRAS, 176, 633

\bibitem{Gezari2012}
Gezari, S. 2012, EPJWC, 39,3001

\bibitem{Ghosh2006}
Ghosh, K. K., Suleymanov, V. et al. 2006, MNRAS, 371, 1587

\bibitem{Greene2007}
Greene, J. E. \& Ho, L. C. 2007, ApJ, 670, 92

\bibitem{Guillochon2013}
Guillochon, J. \& Ramirez-Ruiz, E. 2013, ApJ, 767, 25

\bibitem{Guillochon2015a}
Guillochon, J. \& Ramirez-Ruiz, E. 2015a, ApJ, 798, 64

\bibitem{Guillochon2015b}
Guillochon, J. \& Ramirez-Ruiz, E. 2015b, ApJ, 809, 166

\bibitem{Hills1975}
Hills, J. G. 1975, Nature, 254, 295

\bibitem{Hryniewicz2016}
Hryniewicz, K. \& Walter, R. 2016, A\&A, 586, A9

\bibitem{Ho2008}
Ho, L. C. 2008, ARA\&A, 46, 475

\bibitem{Irwin2010}
Irwin, J.A., Brink, T. G., Bregman, J. N. \& Roberts, T. P. 2010, ApJ, 712, L1

\bibitem{Jonker2010}
Jonker, P. G., Torres, M. A. P. et al. 2010, MNRAS, 407, 645

\bibitem{Jonker2013}
Jonker, P. G., Glennie, A. et al. 2013, ApJ, 779, 14

\bibitem{Kann2011}
Kann, D. A., Greiner, J. \& Rau, A. 2011, ATEL \symbol{35}3813

\bibitem{Kesden2012}
Kesden, M. 2012, Phys. Rev. D, 85, 024037

\bibitem{Komossa2012}
Komossa, S. 2012, EPJWC, 39, 2001

\bibitem{Komossa2015}
Komossa, S. 2015, Journal of High-Energy Astrophysics, 7, 148

\bibitem{Kormendy1995}
Kormendy, J., \& Richstone, D. 1995, ARA\&A, 33, 581

\bibitem{Kormendy2001}
Kormendy, J. \& Gebhardt, K. 2001, in Wheeler J. C., Martel, H., eds, AIP Conf. Proc. Vol. 586, 20th Texas Symposium On Relativistic Astrophysics. Am. Inst. Phys., New York, p. 363

\bibitem{Krolik2011}
Krolik, J. H. \& Piran, T. 2011, ApJ, 743, 134

\bibitem{Lacy1982}
Lacy, J. H., Townes, C. H. \& Hollenbach, D. J. 1982, ApJ, 262, 120

\bibitem{Latif2013}
Latif, M. A., Schleicher, D. R. G. et al. 2013, MNRAS, 433, 1607

\bibitem{Li2002}
Li, -X. L., Ramesh, N. \& Kristen, M. 2002, ApJ, 576, 753

\bibitem{Lodato2009}
Lodato, G., King, A. R., Pringle, J. E. 2009, MNRAS, 392, 332

\bibitem{Lodato2010}
Lodato, G. \& Rossi, E. M. 2010, MNRAS, 410, 359

\bibitem{MacLeod2012}
Macleod, M., Guillochon, J. \& Ramirez-Ruiz, E. 2012, ApJ, 757, 134

\bibitem{Madau2001}
Madau, P. \& Rees, M. J. 2001, ApJ, 551, L27

\bibitem{Maksym2013}
Maksym, W. P., Ulmer, M. P., Eracleous, M. C., Guennou, L., Ho, L. C. 2013, MNRAS, 435, 1904

\bibitem{Maksym2014a}
Maksym, W. P., Ulmer, M. P. et al. 2014a, MNRAS, 444, 866

\bibitem{Maksym2014b}
Maksym, W. P., Lin, D., Irwin, J. A. 2014b, ApJ, 792, L29

\bibitem{Marconi2003}
Marconi, A. \& Hunt, L. 2003, ApJ, 589, L21

\bibitem{Merloni2013}
Merloni, A. \& Heinz, S. 2013, in Planets, Stars and Stellar Systems, Vol. 6, Extragalactic Astronomy and Cosmology, ed. W. Keel, 503

\bibitem{Miller2002}
Miller, M. C. \& Hamilton, D. P. 2002, MNRAS, 330, 232

\bibitem{Nelson2000}
Nelson, C. H. 2000, ApJ, 544, 91

\bibitem{Phinney1989}
Phinney, E. S. 1989, in Proc. 136th IAU Symp. The Center of the Galaxy

\bibitem{Portegies2002}
Portegies Zwart, S. F. \& McMillan, S. L. W. 2002, ApJ, 576, 899

\bibitem{Ptak1999}
Ptak, A. \& Griffiths, R. 1999, ApJ, 517, L85

\bibitem{Read2011a}
Read, A. M., Saxton, R. D. \& Esquej, P. 2011a, ATEL \symbol{35}3811

\bibitem{Read2011b}
Read, A. M., Saxton, R. D. \& Esquej, P. 2011a, ATEL \symbol{35}3821

\bibitem{Rees1988}
Rees, M., J. 1988, Nature, 333, 523

\bibitem{Sani2010}
Sani, E., Lutz et al.  2010, MNRAS, 403, 1246

\bibitem{Saxton2008}
Saxton R. D., Read A. M. et al. 2008, A\&A, 480, 611

\bibitem{Schlegel1998}
Schlegel, D. J., Finkbeiner, D. P. \& Davis, M. 1998, ApJ, 500, 525

\bibitem{Shiokawa}
Shiokawa, H., Krolik, J. H. et al. 2015, ApJ, 804, 85

\bibitem{Smartt2015}
Smartt, S. J., Valenti, S. et al. 2015, A\&A, 579, A40

\bibitem{Soltan1982}
Soltan, A. 1982, MNRAS 200, 115

\bibitem{Strubbe2009}
Strubbe, L. E. \& Quataert, E. 2009, MNRAS, 400, 2070

\bibitem{Struder2001}
Str$\rm \ddot u$der, L., Briel, U. et al. 2001, A\&A, 365, L18

\bibitem{Turner2001}
Turner, M. J. L., Abbey, A. et al. 2001, A\&A, 365, L27

\bibitem{Volonteri2010}
Volonteri, M. 2010, Nature, 466, 1049

\bibitem{Warner2002}
Warner, B. 2002, AIP Conf. Ser., 637, 3

\bibitem{Wolter2006}
Wolter, A., Trinchieri, G. \& Colpi, M. 2006, MNRAS, 373, 1627

\bibitem{Yu2002}
Yu, Q. \& Tremaine, S. 2002, MNRAS, 335, 965

\end{thebibliography}
\end{document}